\documentclass{article}

\usepackage[nonatbib,preprint]{neurips_2020}
\usepackage[numbers]{natbib}

\usepackage[utf8]{inputenc} 
\usepackage[T1]{fontenc}    
\usepackage{hyperref}       
\usepackage{url}            
\usepackage{booktabs}       
\usepackage{amsfonts}       
\usepackage{nicefrac}       
\usepackage{microtype}      
\usepackage{xcolor}
\usepackage{xspace}
\usepackage{tabularx}
\usepackage{amsmath,bm,multicol,multirow}
\usepackage{cleveref}
\usepackage[pdftex]{graphicx}
\usepackage{array, makecell}
\usepackage[scientific-notation=true]{siunitx}
\usepackage{subcaption}
\usepackage{chngpage}
\usepackage{float}

\usepackage{ulem}
\title{\LARGE Wav2Vec2.0 on the Edge: Performance Evaluation}

\author{%
  Santosh Gondi \\
  Meta Inc
}

\begin{document}

\maketitle

\begin{abstract}
Wav2Vec2.0~\cite{baevski2020wav2vec} is a state-of-the-art model which learns speech representations through unlabeled speech data, aka, self supervised learning. The pretrained  model is then fine tuned on small amounts of labeled data to use it for speech-to-text and machine translation tasks. Wav2Vec 2.0 is a transformative solution for low resource languages as it is mainly developed using unlabeled audio data. Getting large amounts of labeled data is resource intensive and especially challenging to do for low resource languages such as Swahilli, Tatar, etc. Furthermore, Wav2Vec2.0 word-error-rate(WER) matches or surpasses the very recent supervised learning algorithms while using 100x~\cite{baevski2020wav2vec} less labeled data. Given its importance and enormous potential in enabling speech based tasks on world's 7000 languages~\cite{lang7000}, it is key to evaluate the accuracy, latency and efficiency of this model on low resource and low power edge devices and investigate the feasibility of using it in such devices for private, secure and reliable speech based tasks. On-device speech tasks preclude sending audio data to the server hence inherently providing privacy, reduced latency and enhanced reliability. In this paper, Wav2Vec2.0 model's accuracy and latency has been evaluated on Raspberry Pi along with the KenLM~\cite{Heafield11kenlm} language model for speech recognition tasks. How to tune certain parameters to achieve desired level of WER rate and latency while meeting the CPU, memory and energy budgets of the product has been discussed.
\end{abstract}

\section{Introduction}
Deep learning based speech recognition has come a long way in the past decade~\cite{6296526}. DL is a subfield of machine learning which imitates the neural network structure of the human brain for pattern matching and classification. The major reason for its popularity is that it does not need feature engineering. It independently extracts the features based on the patterns in the labeled training dataset. However, it needs large quantities of labeled data to achieve higher prediction accuracy. On the contrary, Self-supervised learning (SSL)~\cite{ssl} doesn’t need labeled data. It learns to understand the structures and patterns of input just based on raw input data. The general technique of self-supervised learning is to predict any unobserved or hidden part (or property) of the input from any observed or unhidden part of the input. This learning is very helpful in cases where labeled data is very difficult to get or doesn’t exist such as for translation or recognition of low resource languages. Wav2Vec2.0, just referred to as Wav2Vec in the rest of the paper, is trained with a self-supervised algorithm using unlabeled LibriSpeech dataset. It learns meaningful representations from raw speech data. Then the model is fine tuned with a small amount of labeled data to use for downstream tasks such as translation and speech recognition. 

Automatic speech recognition has many practical applications on all the digital devices people use everyday. Voice search for music or document, voice to text for messaging and translation, voice assistant for navigating apps or controlling home devices, to name a few. On-device speech recognition has additional benefits of being more secure and reliable by not having to send audio data to servers over a network connection. On-device based speech recognition can enable military, healthcare, voice based accessibility apps where the network connection may not exist. Given the transformative capability, the Wav2Vec is bringing to world languages, the latency and feasibility of it for on-device execution is explored using Raspberry Pi as a proxy for low end edge devices.

The Wav2Vec model outputs the sequence of token probabilities in the alphabet set. A simple arg-max followed by tokenizer yields sufficiently good accuracy. However, by using a language on top of the Wav2Vec output significantly improves the WER, though the decoding using language model increases the memory footprint of the transcription task. In this paper, WER, latency and memory are analyzed using a simple tokenizer as well as the Trie based KenLM language model.

In this paper, the performance of base Wav2Vec models is evaluated on Raspberry Pi and different metrics like accuracy, latency and efficiency trade-offs to consider during deployment are discussed.  This study is an extension of the previous work~\cite{electronics10212697,su132212392}. Here the focus is on Wav2Vec decoding with language model and energy measurements along with accuracy and performance.  The previous ASR evaluations on edge devices have focused on RNN~\cite{graves2012sequence}/LSTM~\cite{lstm,lstmtut} based models\cite{9208978}. The transformer~\cite{vaswani2017attention} models are based on different architecture and boost the accuracy quite significantly. There is no other work which evaluated the latency and feasibility of using latest ASR systems such as Wav2Vec, on the edge devices.

 The rest of the paper is organized as follows: In the background section Wav2Vec and language models are discussed. Experimental setup section covers the steps for preparing the model and setting up for inferencing and energy measurements. Accuracy, latency and efficiency metrics are covered in the results section. Finally, paper ends with a summary and outlook.

\section{Background}
Wav2Vec takes raw 16K sampled audio as input. It encodes speech audio through multi-layer CNN and then resulting speech representations are masked before being fed to a Transformer network. The model is trained via a contrastive task where the true latent representations are to be distinguished from distractors. Transformer is a sequence-to-sequence architecture originally proposed for machine translation~\cite{vaswani2017attention}. It has since been adopted for ASR, with the difference being that the input is audio frames instead of the text like in translation tasks. The Wav2Vec feature encoder contains seven blocks and the temporal convolutions in each block have 512 channels with strides (5,2,2,2,2,2,2) and kernel widths (10,3,3,3,3,2,2)~\cite{baevski2020wav2vec}. Base model contains 12 transformer blocks, model dimension 768, inner dimension (FFN) 3,072 and 8 attention heads. 

Wav2Vec inference with simple arg-max tokenizer(greedy) as well as using beam search decoding with KenLM language models is explored. Beam search explores the search space in a greedy breadth-first fashion retaining only certain top candidates in each step~\cite{vijayakumar2018diverse}. The number of candidates kept during search is called beam width. Beam search with a language model is used to boost the accuracy of speech-to-text output. The emitted logits with probabilities over alphabet space are overlapped with language model probabilities for potential sentences, keeping only few top candidates at each step. Using higher beam sizes might yield better accuracy but is not practical to use very high beam size on memory constrained devices. Experiments in this paper have explored the beam size of 10 and 100 and analysis of the tradeoff between accuracy and memory is provided.

KenLM~\cite{Heafield11kenlm} is a widely used library that implements data structures for efficient language model queries. Trie based KenLM for decoding is used has been used for experiments in this paper . It uses Trie data structure with bit-level packing, sorted records, interpolation search for efficient memory usage. A 4-gram KenLM language model for the LibriSpeech dataset is used in the experiments. 

Figure 2 shows the simplified flow of the ASR process with Wav2Vec model.
 
\begin{figure}[H]
\centering
\includegraphics[width=0.9\textwidth]{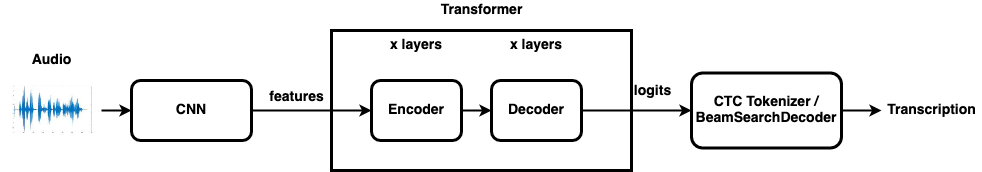}
\caption{Wav2Vec inference.}
\label{fig3}
\end{figure}

\section{Experimental Setup}
\subsection{Model preparation}
PyTorch\cite{pytorch} based models are used for the evaluation. PyTorch is an open source machine learning framework based on torch library. Figure 3 shows the steps for preparing the models for inferencing on edge devices.

\begin{figure}[H]
\includegraphics[width=0.9\textwidth]{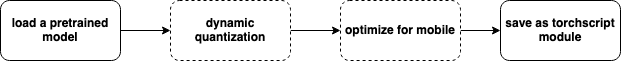}
\caption{Model preparation steps~\cite{su132212392}}
\label{fig3}
\end{figure} 

This section goes over a few of the PyTorch tools and APIs used in this evaluation.

\subsubsection{TorchScript} TorchScript is the means by which PyTorch models are productionized. The python based modules are  optimized, serialized and saved in intermediate representation(IR) format when converted to a torch script module for mobile inference~\cite{torchjit}. TorchScript modules are created by tracing or scripting methods. Tracing works by executing the model with sample inputs and capturing the computations. Whereas, scripting does static inspection by going through the model recursively.  The benefit of scripting is that it correctly handles the loops and control statements in the module. A saved script module can then be loaded either in a Python or C++ environment for inferencing purposes. ScriptModules for Wav2Vec is generated by scripting method after applying quantization and optimizations. The inferencing is done using Python. 

\subsubsection{PyTorch mobile} PyTorch mobile~\cite{ptmobile} provides a set of APIs for optimizing the models for mobile platforms. It uses module fusing, operator fusing, quantization among other things to optimize the models. By quantization, model weights and computation are reduced to int8 precision from float precision. This reduces memory, storage and computation cycles of the model. The device backend was set to qnnpack during inference in the script as : \emph{torch.backends.quantized.engine='qnnpack'}. QNNPACK\cite{qnnpack} (Quantized Neural Networks PACKage) is a mobile-optimized library for low-precision high-performance tensor computations.

\subsubsection{Wav2Vec model}
A Google Colab instance was used for importing, optimizing and saving the model as TorchScript module. Pretrained Wav2Vec models from \emph{huggingface} model hub\cite{hfwav} were used. For this evaluation, a base model with 100hr and full 960hr hour fine-tuning was used. 100hr fine tuned means the base Wav2Vec model was trained on 960hr of LibriSpeech unlabeled audio data with self-supervised learning. This model would have learned the basic speech structures. Then a 100hr labeled audio data is used to fine tune the model with CTC\cite{1143891} loss function for speech recognition task in english. The base model has 93 million parameters. The unquantized original model is scripted format is 377 MB. where as the quantized scripted model is 207 MB. The model dimentionality doesn't vary between 100hr and 960hr fine tuned models. The Wav2Vec model emits the logits(probability of tokens). The decoding process which coverts these token probabilities to text transcriptions could be simple tokenizer or a BeamSearch decoder with a language model. In this experiment the accuracy and latency with both tokenizer and language model based decoding was evaluated.

\subsubsection{KenLM}
BeamSearchDecoderCTC module from pyctcdecode\cite{pyctc} for beam search decoding was used. A readily available 4-gram librespeech language model from~\cite{lslm} was used.  KenLM repository was built from source to generate binary executables, such as \emph{build\_binary}, required for converting 4-gram ARPA language models to Trie based binary files. The same utility was also used for quantizing the language model. ARPA language model is text based, large in size and takes longer to load. It is good for interoperability between different packages. However, in production usecases it is converted to binary format to reduce loading latency and memory footprint.

\subsection{Datasets}
LibriSpeech~\cite{7178964}, a standard  ASR  bench-marking dataset with $\sim$1,000 hours of labeled english speech from various audiobooks was used for evaluation. It has a combination of train, dev and test datasets. For evaluation, test and dev data sets are used. 

\subsection{Raspberry Pi setup}
Raspberry Pi 4 B is used for evaluation. It is a basic headless module( Figure \ref{fig:energy_c}) with no additional peripherals attached to it. Relevant device specs are provided in Table \ref{tbl:rbpispec}. The default Raspberry Pi OS is 32 bit which is not compatible with PyTorch. Therefore, a 64 bit OS version was installed on Pi.

\begin{table}[H]
\caption{Raspberry Pi 4 B spec.\label{tbl:rbpispec}}
\centering
\begin{tabular}{ll}
\toprule
\textbf{Name}	& \textbf{Spec}\\
\midrule
Chip		& BCM2711\\
CPU		& Quad core Cortex-A72 (ARM v8) 64-bit SoC\\
Clock speed		& 1.5GHz\\
RAM		& 4GB SDRAM\\
Caches      & 32 KB data + 48 KB instruction L1 cache per core. 1MB L2 cache\\
Storage		& 32GB micro SD card\\
OS		& 64bit Raspberry Pi OS\\
pip packages & torch, transformers, pyctcdecode, numpy, jiwer, soundfile\\
Python version		& 3.7\\
Power supply		& 5V DC via USB-C connector \\
\bottomrule
\end{tabular}
\end{table}

\subsubsection{Core and Thread configurations}
The ASR inference was studied by configuring the Python script to run on 1, 2, 3 and 4 CPU cores to analyze tradeoffs in terms of accuracy, latency and efficiency. Linux  \emph{taskset} \cite{taskset} command is used to associate evaluation script to certain CPU cores. Using different intra-op and inter-op thread settings in PyTorch\cite{ptthread} was also explored to examine the similar tradeoffs. Experiments showed that the inter-op thread settings do not have an effect on latency and intra-op thread settings have similar effects as core allocations reserved using linux based taskset command.

\subsection{Energy measurements}
\begin{figure}[H]
\begin{minipage}{0.5\textwidth}
\includegraphics[width=0.9\textwidth]{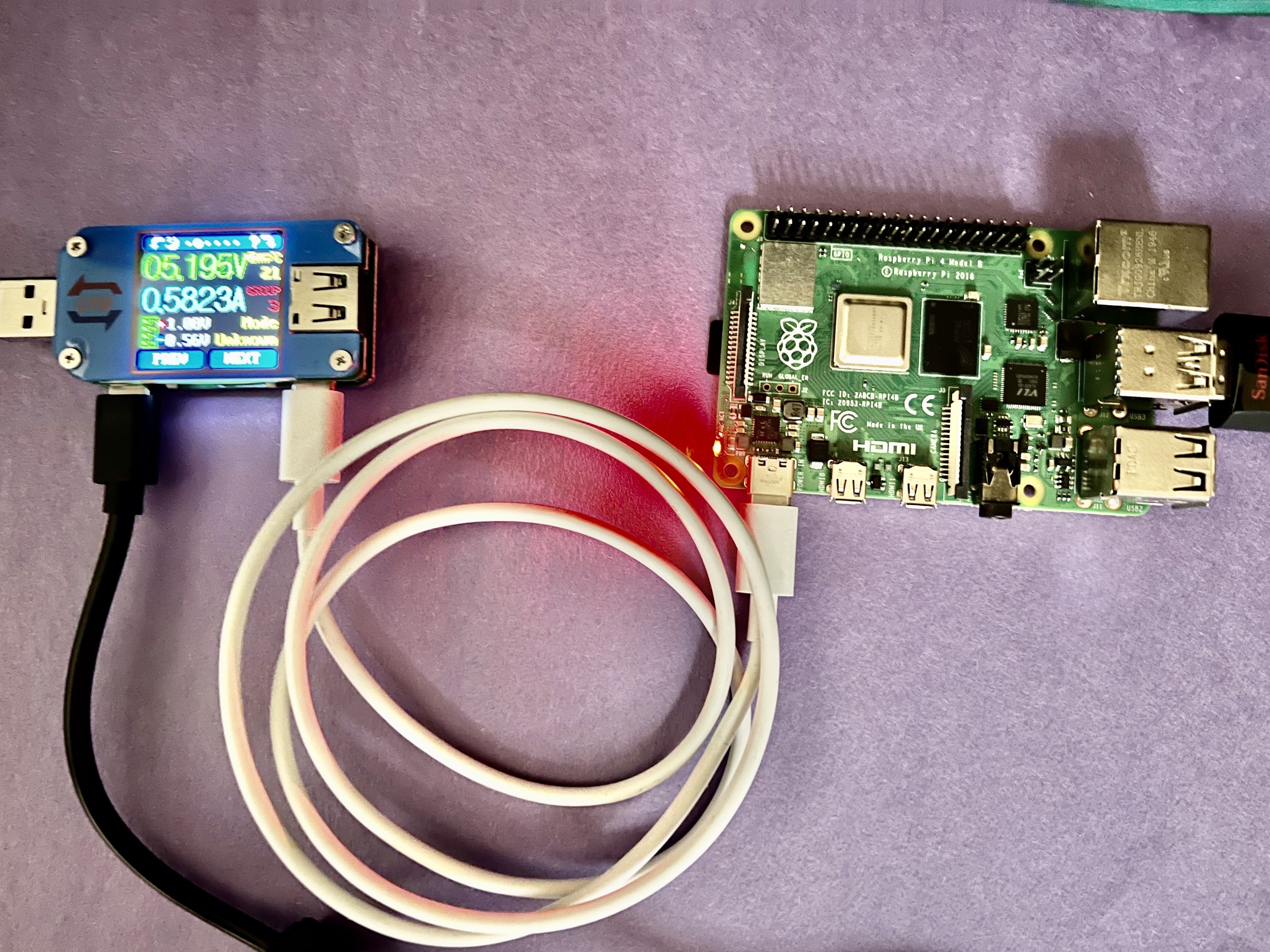}
\caption{Connection setup\label{fig:energy_c}}
\end{minipage}\hfill
\begin{minipage}{0.5\textwidth}
\includegraphics[width=0.9\textwidth]{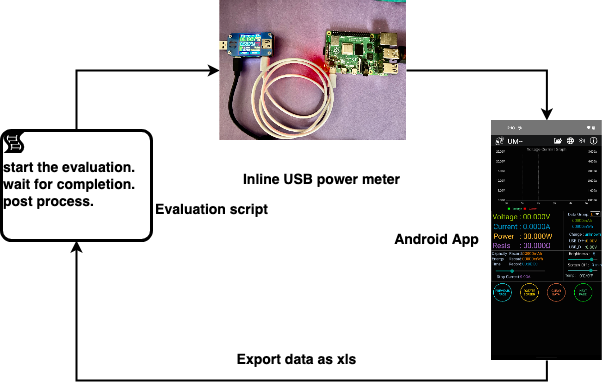}
\caption{Measurement flow \label{fig:energy_s}}
\end{minipage}
\end{figure}

A USB power meter\cite{Makerhawk} for measuring the power consumption on Raspberry Pi was used. Figure \ref{fig:energy_c} shows the connection setup and  Figure \ref{fig:energy_s} shows the flow for monitoring energy during ASR inference. The monitoring device provides voltage and current values every second. Watts and mWh(milli-watt-hours) during different states of the experiment were computed from these continuous voltage and current values. An app on the android phone connects to USB meter over bluetooth and collects voltage and current values during experiment and the readings are saved in a file. The execution script reads the data from file and computes the energy/power usage during specific configuration of experiment.

\section{Results}
In this section accuracy(WER), latency(RTF) and efficiency(CPU, memory and power consumption) metrics are presented.

LibriSpeech test and dev datasets together contain $\sim21$ hours of labeled audio data. It would have taken many days for the inference to execute on all combination of models with all the audio in these data sets. To save time, 10\% of the audio files in each of the four data sets were sampled for inference and evaluation. The same sampled datasets were used in all the test runs thus ensuring that the results are comparable.

\subsection{WER}
Table \ref{tbl:wer} shows the WER of quantized and unquantized models for all the datasets with different LM configurations. WER reported in ~\cite{baevski2020wav2vec} is also shown as Server with LM for comparison.

\begin{table}[h] 
\caption{WER(\%)\label{tbl:wer}}
\centering
\begin{tabular}{ llcccccc }
\toprule
\textbf{Dataset}  & \textbf{Model} &\textbf{No LM} & \multicolumn{2}{c}{\textbf{full LM}}  & \multicolumn{2}{c}{\textbf{quantized LM}}  & \textbf{Server with LM}\\
\cmidrule(r){4-5}\cmidrule(l){6-7}
& & & \textbf{b\_w 10} & \textbf{b\_w 100} & \textbf{b\_w 10} & \textbf{b\_w 100}& \\
\midrule
\multirow{4}{*}{test-clean} & 100hr quantized	& 6.8	 & 4.8	 & 4.6 	 & 5.1	 & 4.8	 & \\
& 100hr	& 6.4 & 4.9	 & 4.6 	 & 4.8	 & 4.6	 & 3.4\\
& 960hr quantized	& 4.0	 & 3.3	 & 3.3 	 & 3.3	 & 3.4	 & \\
& 960hr		& 3.7	 & 3.1	 & 3.1 	 & 3.1	 & 3.1	 & 2.6\\
\midrule
\multirow{4}{*}{test-other} & 100hr quantized	& 15.5	 & 11.2	 & 10.4 	 & 11.5	 & 10.4	 & \\
& 100hr	& 14.9	 & 11.2	 & 11.0 	 & 11.2	 & 11.0	 & 8.0\\
& 960hr quantized	& 10.0	 & 7.5	 & 7.2 	 & 7.5	 & 7.2	 & \\
& 960hr		& 9.0	 & 7.6	 & 7.4 	 & 7.6	 & 7.4	 & 6.1\\
\midrule
\multirow{4}{*}{dev-clean} & 100hr quantized	& 7.8 & 4.7	 & 4.3 	 & 4.7	 & 4.3	 & \\
& 100hr	&7.0  & 4.4	 & 4.1 	 & 4.4	 & 4.1	 & 2.7\\
& 960hr quantized	& 4.7	 & 3.3	 & 3.2 	 & 3.2	 & 3.1	 & \\
& 960hr		& 4.2	 & 3.2	 & 3.2 	 & 3.2	 & 3.2	 & 2.0\\
\midrule
\multirow{4}{*}{dev-other} & 100hr quantized	& 15.8 & 11.0	 & 9.9 	 & 11.0	 & 9.8	 & \\
& 100hr	& 15.0	 & 10.7	 & 9.8 	 & 10.6	 & 9.7	 & 7.9\\
& 960hr quantized	& 11.0	 & 8.7	 & 8.1 	 & 8.7	 & 8.2	 & \\
& 960hr		& 10.3	 & 7.8	 & 7.5 	 & 7.8	 & 7.5	 & 5.9\\
\bottomrule
\end{tabular}
\end{table}

Measurements are done with 100-hour and 960-hour pretrained models. With the language model, the WER is at least  $\sim$30 pcs better than the WER measured for inference without language model.  Furthermore, there is at least $\sim$3 pcs improvement with the beam size of 100 compared to beam size of 10.  With the quantized language model, the WER is $\sim$25 pcs better than inference run without language model. 
Between quantized and unquantized models, the unquantized language model yields the best error rate. However the quantized model has lower memory footprint which may be desirable for low resource edge devices such as smart speakers, displays, thermostats, etc. The other datasets show higher WER than clean data sets. This is expected since the other data set is noisier than clean datasets. Also, server side WER is lower in all the cases because in the paper~\cite{baevski2020wav2vec} unquantized Wav2Vec and language model with 1500 beam size was used. On low resource edge devices, it is not feasible to use very high beam size because memory won't be sufficient to support it. Considering the memory constraints of embedded environments, experiments with beam widths of 10 and 100 were considered. Beam width 50 was also tested which yielded similar results as 100. 

\subsection{RTF}
RTF(real-time-factor) is defined as inference time divided by utterance duration. For example, a RTF of 0.5 means a 5 sec utterance can be transcribed in 2.5 secs. Table \ref{tbl:rtf_normal} shows the RTF of quantized and unquantized models with different core and thread configurations. Since RTF doesn't vary across datasets, just the dev-clean data set was used for RTF evaluations. Different set of CPU cores were configured to simulate environments similar to embedded devices like wearables and hand held devices. WER remained the same in different core and thread configurations. However, RTF increased as the number of cores allowed to be used by PyTorch reduced. The key finding from below experiments was that the RTF didn’t degrade much between 3 and 4 cores. Another experiment was done where the inference was allowed to run on all the cores (4 cores) but the number of threads that PyTorch can use for intra-op parallelization was controlled. The default intra-op threads used by PyTorch is 4. An interesting finding is when 2 threads are configured and allowed to run on 4 cores, the RTF was very similar to using the default of 4 threads. Using language model for decoding didn't noticeably increase the RTF.

\begin{table}[!htb]
    \caption{RTF\label{tbl:rtf_normal}}
    \begin{minipage}{.55\linewidth}
      \subcaption{Core configs}
      \centering
        \begin{tabular}{lcccc}
             \toprule
\textbf{Model}  & \textbf{1 core} & \textbf{2 cores} & \textbf{3 cores} & \textbf{4 cores}\\
\midrule
W2V quant		& 2.3 & 1.35 & 1.07 & 0.98 \\
W2V 	& 2.48 & 1.66 & 1.45  & 1.36 \\
 \bottomrule
        \end{tabular}
    \end{minipage}%
    \begin{minipage}{.45\linewidth}
      \centering
        \subcaption{Intra-op thread configs}
        \begin{tabular}{lcc}
            \toprule
\textbf{Model}  & \textbf{1 thread} & \textbf{2 threads}\\
\midrule
W2V quant		& 1.51 & 1.13\\
W2V 	& 1.70 & 1.42\\
 \bottomrule
        \end{tabular}
    \end{minipage} 
\end{table}

\subsection{Energy}
\begin{figure}[H]
\begin{minipage}{0.5\textwidth}
\includegraphics[width=0.9\textwidth]{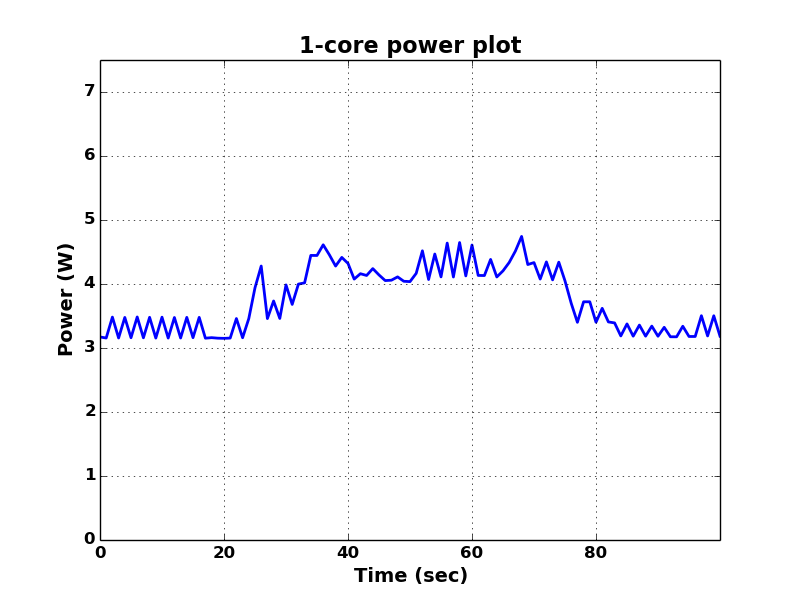}
\caption{Power plot with 1 core\label{fig:power1}}
\end{minipage}\hfill
    \begin{minipage}{0.5\textwidth}
\includegraphics[width=0.9\textwidth]{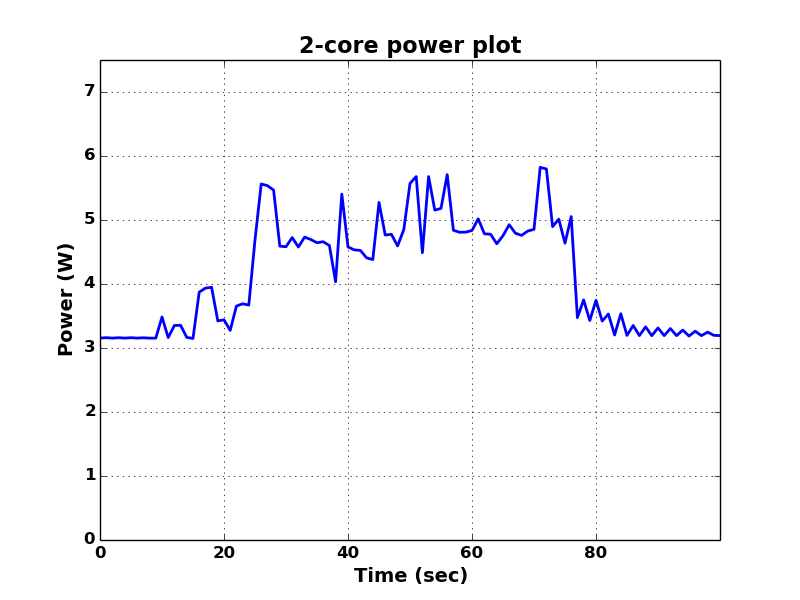}
\caption{Power plot with 2 core \label{fig:power2}}
\end{minipage}
\end{figure}

\begin{figure}[H]
\begin{minipage}{0.5\textwidth}
\includegraphics[width=0.9\textwidth]{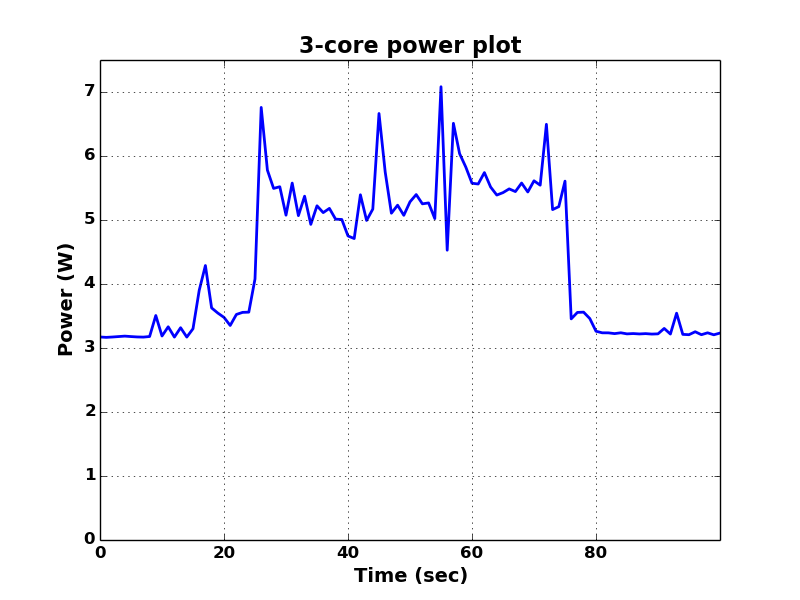}
\caption{Power plot with 3 core\label{fig:power3}}
\end{minipage}\hfill
    \begin{minipage}{0.5\textwidth}
\includegraphics[width=0.9\textwidth]{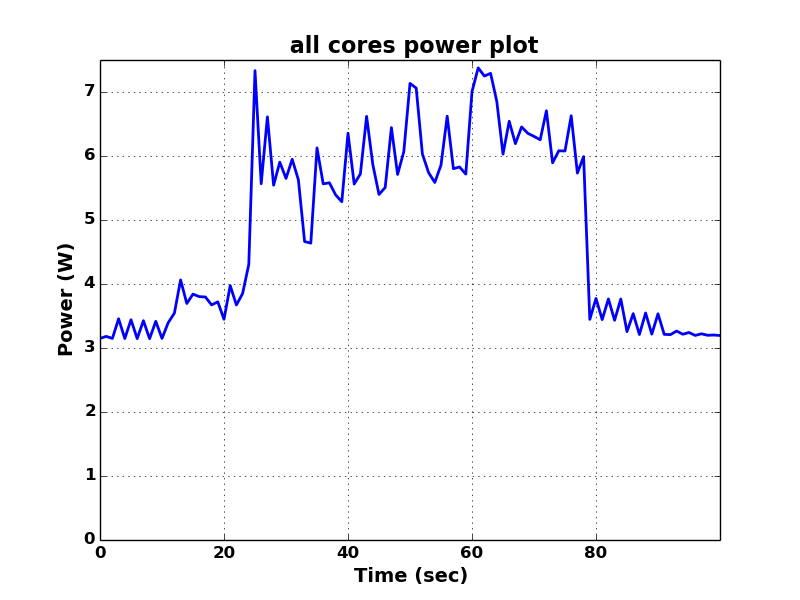}
\caption{Power plot with all cores \label{fig:power4}}
\end{minipage}
\end{figure}

Figure \ref{fig:power1}, \ref{fig:power2} , \ref{fig:power3} and \ref{fig:power4} show the power variation when quantized model is run with 1, 2, 3 and 4 core environments.  From the graph, it can be seen that the energy consumption is higher as the number of cores increase. The steady state wattage in normal mode is $\sim3.1 W$ (5.2 V x 0.6 A) on Raspberry Pi. This increases by $\sim1.1 W$, $\sim1.7 W$, $\sim2.3 W$ and $\sim2.9 W$ for 1,2,3 and 4 cores, respectively during ASR inference. 

\begin{figure}[h]
\begin{minipage}{0.5\textwidth}
\includegraphics[width=0.9\textwidth]{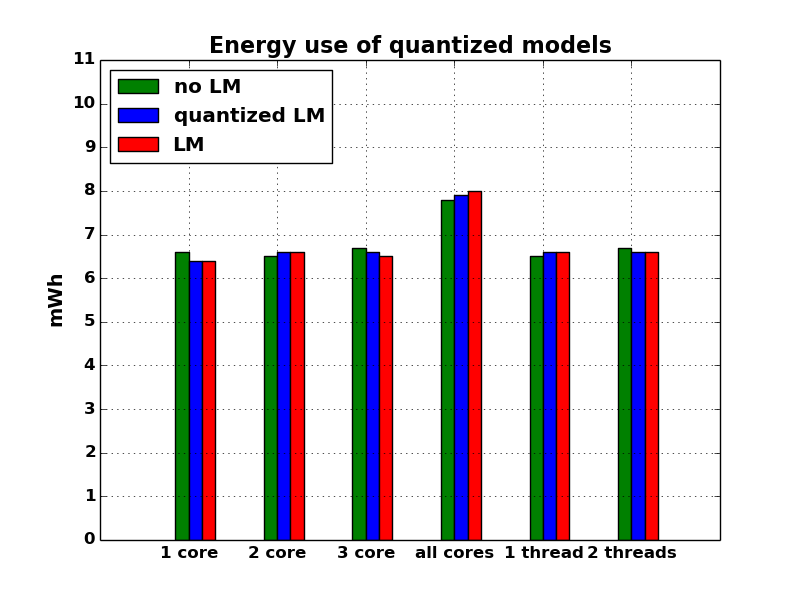}
\caption{For quantized models \label{fig:energy_bar_q}}
\end{minipage}\hfill
    \begin{minipage}{0.5\textwidth}
\includegraphics[width=0.9\textwidth]{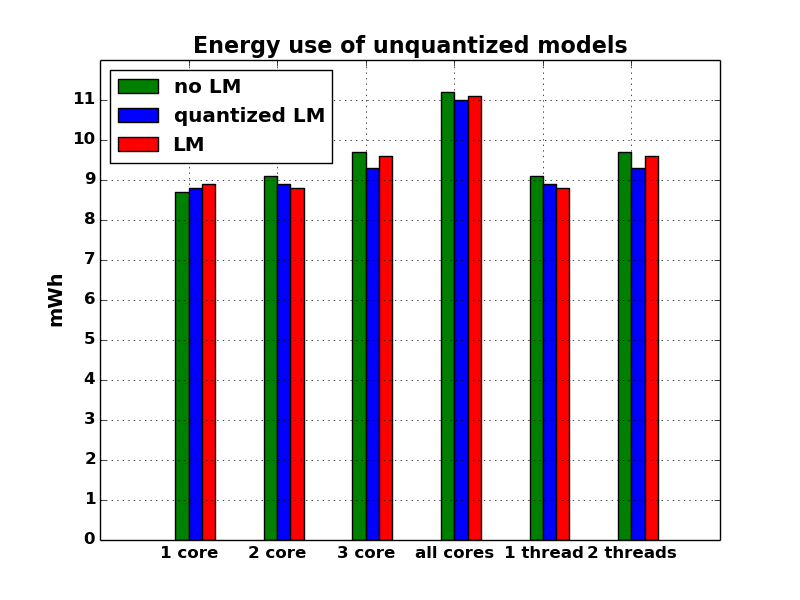}
\caption{For unquantized models \label{fig:energy_bar}}
\end{minipage}
\end{figure}

Figures \ref{fig:energy_bar_q} \& \ref{fig:energy_bar}  shows the energy consumption of quantized and unquantized models with different core and thread configurations for converting 10 second long utterance to text. The unquantized full model consumes $\sim11mWh$ of energy for 10 secs of utterance whereas the quantized model consumes $\sim27\%$ less energy compared unquantized model. By selectively configuring fewer cores or using 2 or fewer intra-op threads, $\sim20\%$  energy is saved. Using language model decoding doesn't noticeably increase the energy consumption.

\subsection{Memory and CPU}
\begin{figure}[h]
\begin{minipage}{0.5\textwidth}
\includegraphics[width=0.9\textwidth]{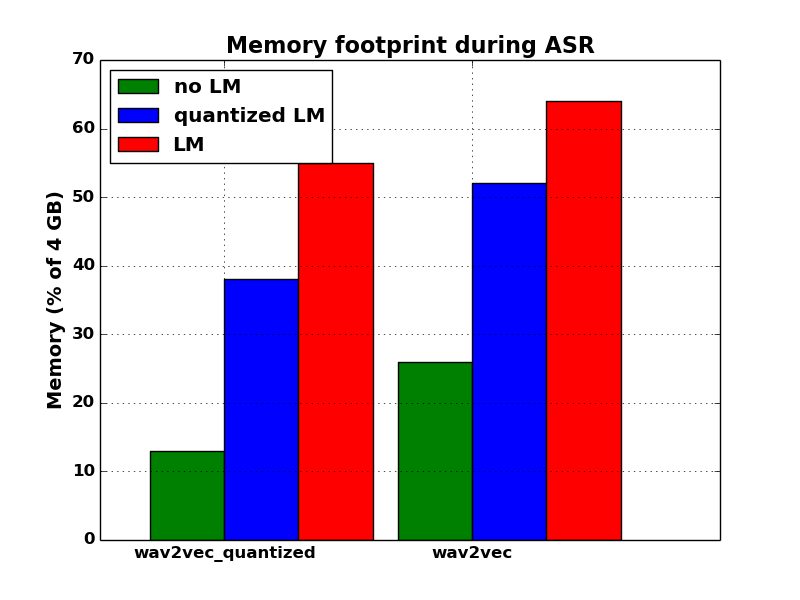}
\caption{Memory footprint\label{fig:mem}}
\end{minipage}\hfill
    \begin{minipage}{0.5\textwidth}
\includegraphics[width=0.9\textwidth]{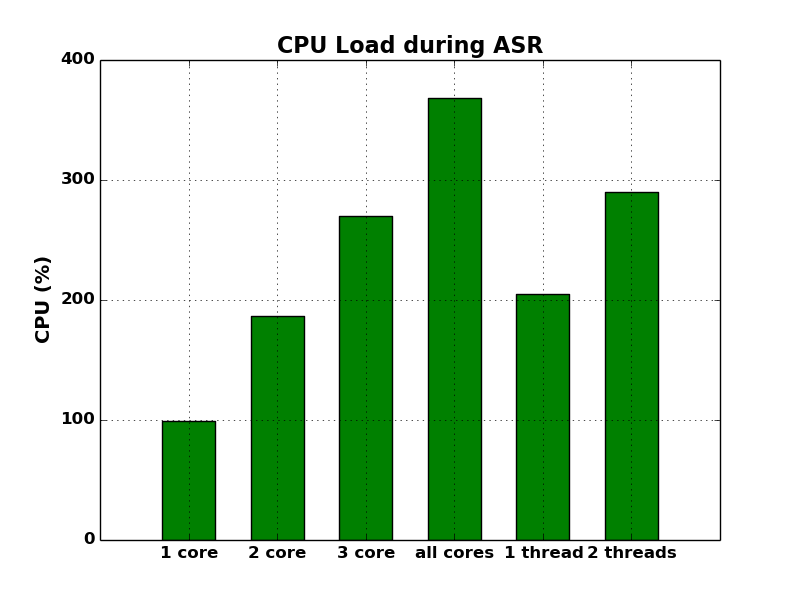}
\caption{CPU load\label{fig:cpu}}
\end{minipage}
\end{figure}

Figures \ref{fig:mem} \& \ref{fig:cpu} shows the memory footprint and CPU utilization during different evaluation setups.  An unquantized language model takes $\sim40$ pcs more memory than quantized version.  Quantized language model on the other hand has a better memory footprint and takes ~25 pcs more memory compared to an inference run with no language model. The configuration of 3 cores vs 2 intra-op threads takes similar amount of CPU.

\section{Discussion}
Different levers can be used to achieve desired accuracy and latency on Wav2Vec while knowing or limiting the memory, CPU and energy costs of the ASR inference with those choices. On a device with fewer CPU cores such as smart thermostat or in multi-user scenarios such as playing the music while sending a dictated message with smart glass, 2 or 3 cores can be allocated for inference while leaving the rest of the cores for other tasks. If the device is less on memory, such as smart watch a quantized model with no language model can be chosen. Lesser memory comes with reduced ASR accuracy but that may be tolerable on device with limited set of voice commands. If there is no memory or CPU resource constraint on the device such as modern smartphone quantized Wav2Vec model with LM support can be chosen to achieve accurate speech recognition for accessibility needs or hands free messaging. On a device with small battery such as a smart watch, the model config with least amount of energy consumption can be chosen after modeling the typical usage of the device in a day and how often the speech recognition is used in a day. For example, if a user sends 10 messages per day using dictation, speech to text conversion with Wav2Vec quantized model would consume $\sim$9\% of battery(considering Apple Watch battery capacity of 309 mWh) for average utterance length of 5 secs.

\section{Conclusion}
Wav2Vec model inference on raspberry Pi with different configurations has been evaluated. The pre-trained and fine tuned model achieves near realtime (RTF=1) latency with state-of-the-art WER on Raspberry Pi. The memory accuracy of the model improves $\sim30\%$ by using language model for decoding albeit with $\sim200\%$ of additional memory cost. By using quantized language model, memory footprint can be reduced substantially with small sacrifice to accuracy. The energy footprint of quantized model is $\sim27\%$ lower than unquantized model. The energy footprint is strongly co-related with CPU usage. Considering these points, it can be concluded that the model isn't suitable for executing on every resource constrained embedded devices because of high memory footprint. But it can be used in an application on modern smartphone for on-device speech recognition. With further quantization, pruning and distillation, resource footprint of Wav2Vec inference can be reduced.

\begin{ack}
I would like to thank Vineel Pratap, a Research Engineer and co-worker in Meta, for discussing the idea of evaluating transformer performance on edge devices.
\end{ack}

\bibliography{main}

\bibliographystyle{abbrvnat}

\end{document}